\documentstyle[12pt]{article}

\begin{document}

\title{Scaling properties of localization length in 1D paired correlated
binary alloys of finite size}

\author{
Felix M. Izrailev$^{1,2}$\thanks{email addresses: izrailev@vxinpb.inp.nsk.su ; 
izrailev@physics.spa.umn.edu},Tsampikos Kottos$^1$ and G.  P. Tsironis$^{1}$\\ 
\\
$^1$ Department of Physics, University of Crete\\
and Research Center of Crete, P.O. Box 1527\\
71110 Heraklion-Crete, Greece\\
\\
$^2$ Budker Institute of Nuclear Physics,Novosibirsk 630090, Russia}

\maketitle

\begin{abstract}
We study scaling properties of the localized eigenstates of the random dimer
model in which pairs of local site energies are assigned at random in a one
dimensional disordered tight-binding model. We use both the transfer matrix
method and the direct diagonalization of the Hamiltonian in order to find
how the localization length of a finite sample scales to the localization
length of the infinite system. We derive the scaling law for the localization
length and show it to be related to scaling behavior typical of uncorrelated
Band Random Matrix, Anderson and Lloyd models.
\end{abstract}

\baselineskip 24 pt \vspace{2in} \newpage

\newpage

\baselineskip 24 pt \vspace{2in}

\newpage


\section{\bf Introduction}

It is well known that in one-dimensional (1D) disordered models even small amount
of disorder leads to an exponential localization of all eigenstates \cite
{A-58,E-79}. On the other hand, recent studies of quasi-1D
polymers have shown that short-range correlations embedded in a random
sequence can lead to appearance of fully transparent states \cite
{DWP-90,PW-91}. In ref. \cite{PW-91} in particular, various organic
disordered systems were quoted with electrical properties. The prototypical
case is that of the Random Dimer Model (RDM) \cite{DWP-90,PW-91} where (in
the context of a tight-binding Hamiltonian) pairs of adjacent energy levels
are assigned at random, leading to two-site correlations in an otherwise
random model.

Since for infinite samples fully delocalized states appear only for specific
energy values, there is no Anderson transition in the usual sense (see also 
\cite{GS-92}). However, the number of transparent states for finite samples
was found to be proportional to square root of the length of the sample \cite
{DWP-90,PW-91}. This fact is related to the divergence of the localization
length in infinite samples when the energy approaches some critical values 
\cite{F-89,B-92,IKT-95}. Therefore, these states may be important for
conducting properties of finite samples \cite{DGK-93,DGK-93a}.

In infinite samples the Anderson transition can be characterized in terms of
the localization length; the latter is commonly defined from the decay of
amplitude of eigenstates in the limit $|n|\rightarrow \infty $, where $n$ is
the site label in the tight-binding picture. Contrary to what happens in
infinite samples, the global properties of eigenstates of finite samples
cannot be characterized in the same way; one needs to use other quantities
(such as the inverse participation ratio), that are valid both for finite
and infinite samples. Then, through the use of scaling conjectures, one can
link the properties of eigenstates in infinite samples to those in finite
samples. In the theory of disordered solids, the scaling approach proved to
be extremely useful in describing the conductance and its fluctuations (see
e.g. \cite{AALR-79,P-86}). A similar approach has been recently used in the
theory of quantum chaos when describing localized eigenstates random on a
finite scale \cite{CGIS-90,I-90}. Such eigenstates also arise in the quasi-1D 
models with random potential. Extensive numerical and analytical studies
(see e.g. \cite{FM-94,I-95} and references therein) have revealed remarkable
scaling properties of eigenstates, which seem to be of quite generic nature.

In this paper we study the RDM of finite size from the point of view of
scaling properties of its eigenstates. The question of the relevance of the
above mentioned results to models with correlated disorder is far from
being trivial since short range correlations may cause significant
difference in the structure of eigenstates, when compared with those for
random potentials. In the next Section 2 we briefly describe the RDM and
discuss different definitions of localization length, which are used in our
numerical simulations. In Section 3 we present numerical data on scaling
properties of eigenstates in the center of energy bands. In this case, the
localization length in infinite samples has been obtained by the transfer
matrix method. In Section 4, we study the energy region near the
critical values $E_{cr}$ by making use both numerical and analytical treatment of the
localization length. Finally, in Section 5 we give a short summary of our
investigation.

\section{\bf Finite size scaling approach to random dimer model}

Our starting point is the 1D Schr\"odinger equation in the
tight-binding approximation, 
\begin{equation}
\label{tbe}i{\frac{dc_n(t)}{dt}}=\,\epsilon _nc_n(t)+c_{n+1}(t)+c_{n-1}(t)\
\,, 
\end{equation}
where $c_n(t)$ is the probability amplitude for an electron to be at site $n$
and $\epsilon _n$ is the local site energy. By making the transformation $%
c_n(t)=\exp (-i\,E\,t)\,x_n$ one can obtain the equation 
\begin{equation}
\label{tbe2}E\varphi _n=\varphi _{n+1}+\epsilon _n\varphi _n+\varphi
_{n-1}\,, 
\end{equation}
for the eigenvalue $E$ and the corresponding eigenstate $\varphi _n(E)$ . In
what follows, we consider the RDM \cite{DWP-90,IKT-95,EW-93} which implies short
range correlations in the sequence of random $\epsilon _n$. In this model
there are only two values of $\epsilon _n$ viz. $\epsilon _A$ or $\epsilon
_B $ that appear in pairs in the sequence $\epsilon _n$'s . In other words,
in order to create the dimer chain the pairs $AA\,$ and $BB$ each with
energies $\epsilon _A$, $\epsilon _B$ respectively are distributed at
random. We take for simplicity the probabilities, of occurrence of the
paired to be equal, i.e. ${\cal P}_{AA}={\cal P}_{BB}=1/2$ .

The RDM has been well studied for an infinite chain (see, e.g. \cite
{DWP-90,GS-92,F-89,B-92,IKT-95,EW-93}). Its basic feature is that for two values of
the energy $E_{cr}=\epsilon _A$ or $\epsilon _B$ its eigenstates are extended. 
In the vicinity of these values and for $\epsilon_{A,B}=|\epsilon _A-\epsilon _B|$
which is less than the critical value $\epsilon_{A,B}^{cr}=2$ , the localization
length $l_\infty $ defined through the exponential decay of the amplitude of
eigenfunction diverges as $l_\infty (E)\sim 1/E^2$ (for the specific value $%
\epsilon_{A,B}=2$ , the singularity law $l_\infty (E)\sim 1/E^{}$ holds instead,
see details in \cite{F-89,B-92,IKT-95}). In spite of the fact that for other
values of $E$ inside the spectrum the localization length is finite, the
influence of nearly-transparent states on the electronic properties of
finite samples is strong (see e.g \cite{DGK-93,DGK-93a}). This is related to
the fact that the number of eigenstates with localization length larger than
the size $N$ of the sample is proportional to $\sqrt{N}$.

Our main interest is in the structure of eigenstates for finite samples,
both in the center of energy band and in the vicinity of the critical energies
$E_{cr}$ where localization length in infinite sample diverges. Unlike the 
more simple case of infinite samples, the meaning of localization length for 
finite samples is not clear. Below, we follow the approach developed in the 
theory of quasi-1D disordered solids which is based on the evaluation of 
multifractal localization lengths (see, e.g. \cite{FM-94}). One of the commonly
used quantities in this approach is the so-called entropic localization length,
defined through to the information entropy ${\cal H\ \ }$of eigenstates, 
\begin{equation}
\label{HN}{\cal H}=-\sum\limits_{n=1}^Nw_n\ln
\,w_n;\,\,\,\,\,\,\,w_n=|\varphi _n^2| 
\end{equation}
where $\varphi _n$ is the $n-$th component of an eigenstate in a given
finite basis. For eigenstates normalized as $\sum_nw_n=1$ , the simplest
case of $\varphi _n=N^{-1/2}$ results to the entropy equal to the maximum
value, viz. ${\cal H}=\ln (N)^{}$ . We define therefore the localization
length $l_N$ as the number of basis states occupied by the eigenstate $%
\varphi _n\,$; the latter is equal to $\exp ({\cal H}_N)$ . Similar
definitions have been used for the first time in ref. \cite{PE-76} where
different characteristics of eigenstates have been discussed in solid state
applications. One can see that in the other limit case of an exponentially
localized state with $\varphi _n=l_\infty ^{-1/2}\exp (-|n-n_0|/l_\infty )$
, the quantity $l_N$ is proportional to $l_\infty $ , viz. $l_N\approx
el_\infty $ (assuming $l_\infty \ll N$ ). One should note that, generically,
the amplitudes $\varphi _n$ fluctuates strongly with $n$ and thus the
coefficient of proportionality between $l_N$ and $l_\infty \,$ depends on
the type of fluctuations.

To study the properties of chaotic states, localized on some scale in the
finite basis in \cite{CGIS-90,I-90} it was suggested to normalize the
localization length $l_N$ in such a way that in the extreme case of fully
extended states the quantity $l_N$ is equal to the size of the basis $N$ .
In such an approach, the entropic localization length $l_N^{(1)}$ is defined
as 
\begin{equation}
\label{l1}l_N^{(1)}=N\exp (<{\cal H}_N>\,-\,{\cal H}_{ref}\,)
\end{equation}
In Eq. (\ref{l1}) the ensemble average $<....>$ is introduced over the
number of eigenstates of the same structure. The normalization factor ${\cal %
H}_{ref}$ has the meaning of an average entropy of the completely extended
random eigenstates in the finite basis, therefore, it can be easily found
analytically \cite{CGIS-90}, 
\begin{equation}
\label{HGOE}{\cal H}_{ref}=\psi (\frac N2+1)-\psi (\frac 32)\approx \ln
\,(\frac N{2.07})
\end{equation}
where $\psi $ is the digamma function and the distribution of components $%
\varphi _n$ is assumed to correspond to the Gaussian Orthogonal Ensemble
(GOE). From (\ref{HGOE}) one can see that for $N\gg 1$ the entropic
localization length of random eigenstates, defined simply as $\exp ({\cal H}%
_{ref}),\,$ is approximately $2.07$ times less than $N$ ; this result is due
to gaussian fluctuations in the components $\varphi _n$ .

Analogously, the whole set of localization lengths $l_N^{(q)}\,$ can be
defined in the following way \cite{FM-94,EE-91}:

\begin{equation}
\label{lq}l_N^{(q)}=\,N\,(\frac{<P_q>}{P_{ref}^{(q)}})^{\frac
1{1-q}}\,\,;\,\,\,\,\,\,\,\,\,\,\,\,q\geq 2 
\end{equation}
where 
\begin{equation}
\label{Hq}P_q=\sum\limits_{n=1}^N(w_n)^q\, 
\end{equation}
and \thinspace $P_{ref}^{(q)}$ \thinspace is the average value of $P_q$ for
the reference ensemble of completely extended states. One should note that
for the particular case $q=2$ the quantity $P_2\,$ is known as the participation
ratio; it is widely used in solid state physics. In the limit case of the
GOE, one can find that $P_{ref}^{(2)}=3/N$ , therefore, the inverse
participation ratio $(P_2)^{-1}$ , which is commonly taken as the definition
of localization length, for random eigenstates is 3 times as less as the
``actual'' length $N$ .

In fact, the above expressions for the localization lengths $l_N^{(q)}$ is
defined through the $2q-$th moments of a distribution of components $\varphi
_n$ of eigenstates; non-normalized to \thinspace $P_{ref}^{(q)}$ quantities (%
\ref{lq}) are well known in the multifractal analysis of wave functions.
Such normalization turns out to be extremely important when establishing
scaling properties of eigenfunctions. Indeed, by normalizing the
localization lengths $l_N^{(q)}$ to the size $N$ of the sample, 
\begin{equation}
\label{betaq}\,\beta _q=\frac{l_N^{(q)}}N 
\end{equation}
one can expect, in the spirit of renormalization theory, that the set of
dimensionless parameters $\beta _q$ is the proper quantity to characterize
generic properties of eigenstates for finite samples. According to the
scaling conjecture in the modern theory of disordered solids, it was assumed 
\cite{CMI-90} that for quasi-1D disordered models described by Band Random
Matrices the quantity $\beta _q$ depends on the scaling parameter $\lambda
\, $ only, which is the ratio of the localization length $l_\infty $ for the
infinite sample, to the size $N\,$of the sample itself. Therefore, the
scaling relation can be written as 
\begin{equation}
\label{scal}\beta _q=f_q(\lambda );\,\,\,\,\,\,\,\,\,\lambda =\frac{l_\infty 
}N 
\end{equation}
Detailed studies, both numerical and analytical, have confirmed this
conjecture for different models like the Kicked Rotator Model and Band
Random Matrices (see, e.g. \cite{I-90,FM-94,I-95} and references therein).
Moreover, the scaling function $f_q(\lambda )$ has also been found.

Our main question is whether the relation of the type (\ref{scal}) is also
valid for our dimer model with the correlated disorder. The first nontrivial
question arises about the reference ensemble for the computation of the
average entropy ${\cal H}_{ref}$ . Indeed, in application to 1D Anderson
type models (see details in \cite{CGIFM-92}) the reference ensemble can not
be chosen as an ensemble of full random matrices, like the GOE. This point
is related to the fact that in the Anderson case fully extended states are
not gaussian random functions but just plane waves which arise for zero
disorder. In the dimer model, the situation is even more complicated due to
strong dependence of the localization length on the energy. However, and
this is our expectation, in spite of the presence of the extended states
at the critical energies, scaling properties of the eigenstates in the dimer
model of finite size $N$ are of generic type discovered for 1D and quasi-1D
disordered models.

For this reason and in the spirit of Ref. \cite{CGIFM-92,M-93}, we define the
normalization factors ${\cal H}_{ref}$ and $P_{ref}^{(q)}$ from the solution
of Eq. (\ref{tbe2}) for the zero disorder, $\epsilon _n=0$ ,

\begin{equation}
\label{eigen}E^k=2\cos \frac{k\pi }{N+1}\,, 
\end{equation}
\begin{equation}
\label{EF}\varphi _n^k=\sqrt{\left( \frac 2{N+1}\right) }\sin \frac{nk\pi }{%
N+1}\,, 
\end{equation}
with $k,n=1,\dots ,N$. The entropy ${\cal H}_{ref}$ of the above
eigenfunctions in the large $N$ limit has the same value for every
eigenvalue $E^k$, i.e., 
\begin{equation}
\label{norm}{\cal H}_{ref}=\ln (2N)-1\,, 
\end{equation}
and correspondingly, 
\begin{equation}
\label{norm2}P_{ref}^{(2)}=\frac 3{2N} 
\end{equation}

\section{\bf Scaling properties of localization lengths in the center of
energy bands}

Since all results depend on the difference $\epsilon_{A,B} =|\epsilon _A-\epsilon
_B| $ but not on the actual values $\epsilon _A$ and $\epsilon _B$
separately, we can set $\epsilon _A=0$ for simplicity. One should stress
that both localization lengths $l_\infty $ and $l_N^{(q)}$ are functions of
the energy $E$ . For this reason, in our numerical experiments we consider
ensembles of states specified by the values of the energy $E$ in a small
window $\Delta E$ and by different realizations of random on-site energies $%
\epsilon _n$. We choose the size of the energy window in such a way that for
every chosen value of $\epsilon _B$ the localization length $l_\infty $ is
approximately constant inside this window (in all the cases $\frac{\Delta
l_\infty }{l_\infty }\leq 0.06$ ). 

In order to study scaling properties of the localized eigenstates we have
used the transfer matrix method for infinite chains as well as the direct
diagonalization of the Hamiltonians that one associated with Eq. (\ref{tbe})
, for finite chains of size $N$. To find the localization length $l_\infty $
we have studied the asymptotic behavior of the random matrix product $\prod 
{\em M}_n$, where {\em M}$_n$ is defined through the relation 
\begin{equation}
\label{M}\xi _{n+1}={\em M}_n\xi _n;\,\,\,\,\,\,\,{\em M}_n=\left( 
\begin{array}{cc}
v_n & -1 \\ 
1 & 0 
\end{array}
\right) ;\,\,\,v_n=E-\epsilon _n\,\, 
\end{equation}
for the vector $\xi _n=(x_n,x_{n-1})\,$ with the matrix ${\em M}_n$ known as
the transfer matrix. Then the localization length $l_\infty $ is the inverse
Lyapunov exponent $\gamma $ ; the latter is evaluated as the exponential
decay rate of an initial vector $\xi _1$ , 
\begin{equation}
\label{gamma}l_\infty ^{-1}=\gamma =\lim _{N\rightarrow \infty }\frac 1N 
\frac{\ln \,\prod\limits_{n=1}^N|{\em M}_n\xi _n|}{|\xi _1|}\,. 
\end{equation}
Although the Lyapunov exponent $\gamma $ for finite $N$ depends on a
particular realization of the disorder, for $N\rightarrow \infty $ it
converges to its mean value. For the above calculations we have used
samples of length $5\times 10^5$ for relatively large values of $\epsilon _B$
and up to $4\times 10^6$ for small values of $\epsilon _B$ .

To reveal scaling properties of localization length for finite samples, we
have computed two localization lengths $l_N^{(1)}$ and $l_N^{(2)}$ according
to the relations discussed in the previous Section, with the normalization
factors ${\cal H}_{ref}$ and $P_{ref}^{(2)}$ in the forms (\ref{norm}) and (%
\ref{norm2}). In the computations of these lengths, the energy window was
taken in the center of the spectrum, around the value $E=\epsilon _B/2$ for $%
\epsilon _B$ equals 2, 1.8, 1.6, 1.2, 1, 0.8, 0.6, 0.4, 0.35 and for the
fixed value of $N$. The width of the windows has been numerically chosen to
provide a small change of localization length inside any of windows. The
values of $\beta _1$ and $\beta _2$ are obtained by the averaging over an
ensemble of random samples of size $N=100\div 800$ for the values of $%
\epsilon _B$ cited previously. As a result, the total numbers of eigenstates
in the energy windows were more than 1000.

All the data have been fitted to the scaling function $\beta _q\,$ found for
quasi-1D disordered models \cite{FM-94}:

\begin{equation}
\label{scalfq}\beta _q=\frac{c_q\lambda }{1+c_q\lambda } 
\end{equation}
In fact, this scaling relation is exact only for $q=2$ , however, for other
cases of small values $q$ , including $q=1$ , it is very close to the
correct one (see details in \cite{FM-94}).

Numerical data reported in Fig.1 give clear evidence of a scaling of the
type (\ref{scalfq}). The fitting parameters $c_q$ are equal to $c_1=2.80$
and $c_2=1.55$ . From this figure one can see that the behaviour of $\beta
_q $ is very different in the two limits of very localized $(\beta _q\ll 1)$
and extended $(\beta _q\approx 1)$ eigenstates. The dependence (\ref{scalfq}%
) has the remarkable property which can be seen in other variables,

\begin{equation}
\label{fit}Y_q=\ln (\frac{\beta _q}{1-\beta _q});\,\,\,\,\,\,\,\,\,X=\ln ( 
\frac{l_\infty }N) 
\end{equation}
which are more convenient when considering the whole region of both very
localized and extended states. Indeed, in these variables the scaling has
extremely simple form, 
\begin{equation}
\label{YX}Y_q=a_q+b_q\,X 
\end{equation}
with $b_q=1$ and $a_q=\ln (c_q)$ . The data for the scaling in variables $%
Y,X $ are presented in Fig.2. The fitting parameters $b_{1,2}$ are found to
be quite close to $1$ i.e. $b_1=1.02$ and $b_2=0.98$, for this reason in the
Fig.2 we put $b_1=b_2=1$. The remarkable result is that the above simple
scaling relation holds in a very large region of the scaling parameter $%
\lambda =l_\infty /N$ . According to the fit to the dependence (\ref{YX}),
the values $a_{1,2}$ are: $a_1=1.05$ and $a_2=0.45$, which gives $\Delta
a_{1,2}=a_1-a_2=0.6$. It is very interesting that these values of $a_{1,2}$
are the same as for common Anderson model \cite{M-93} in the center of the
energy band. This fact is very important in establishing the link between
the RDM and Anderson models of finite size. 

It is of special interest to relate the entropy localization length $%
l_N^{(1)}$ and the localization length $l_N^{(2)}$ associated with the
inverse participation ratio. Their interdependence is shown in Fig.3. We see
that they are approximately equal for very localized and very extended
states. It is also clear that $\beta _2$ is always less than $\beta _1$
since $P_N^{(q)}<P_N^{(q+1)}$, due to the definition of Eq. (\ref{Hq}).
Using the definition of Eq. (\ref{scalfq}) one can find the relation between 
$\beta _1$ and $\beta _2$: 
\begin{equation}
\label{beta12}\beta _2=\frac{c\beta _1}{1+(c-1)\beta _1};\,\,\,\,\,c=\frac{%
c_2}{c_1} 
\end{equation}

\section{\bf Scaling of localization lengths near the critical energy.}

In the previous section we have shown that the scaling law (\ref{scalfq}) ,
found for fully disordered 1D and quasi-1D models, also holds in our dimer
model of finite size when considering eigenstates in the center of energy
bands. In a sense, this property may be expected since far from critical
energies where the localization length diverges, the eigenstates are assumed
to be similar to that known for disordered models. The important question is 
whether this scaling holds for all energies inside the band, in particular, near the
critical energies $E_{cr}=\epsilon _A,\epsilon _B$. Direct numerical computation
of the localization length $l_\infty \,$ through the transfer matrix method
is very difficult in this energy region due to very weak convergence of
Lyapunov exponents. For this reason, we have used the analytical expression
which was derived for $l_\infty $ near the critical energies in an approach
developed in \cite{IKT-95} : 
\begin{equation}
\label{our}l_\infty (E)\approx \frac{2\sin {}^2\mu _0}{Q\delta ^2\cos
{}^2\mu _0};\,\,\,\,\,\,\,\,2\cos \mu _0=E
\end{equation}
Here, the factor $Q$ stands for the probability for the pair $\epsilon
_n=\epsilon _{n+1}=\epsilon _B$ to appear, and $\delta $ is defined by the
relation $E=\epsilon _B-\delta $ $\approx \epsilon _B$ \cite{note}. 
We remind that in our case $Q=1/2$ and $\epsilon _A=0$ has
been assumed for the simplicity. From the above expression (\ref{our}) one
can find that if the value of $\epsilon _B$ is far from the stability border 
$E_B=2$ , and the distance $\Delta =2-\epsilon _{B}$ is large
compared to $\delta =\epsilon _B-E$ , the localization length diverges as 
\begin{equation}
\label{loc2}l_\infty \approx \frac{2\Delta }{Q\delta ^2},\,\,\,\,\,\,\,\,\,%
\delta \ll \Delta \ll 1
\end{equation}
In the other limit case of $\epsilon _b=2$ we have 
\begin{equation}
\label{loc1}l_\infty \approx \frac 2{Q\delta },\,\,\,\,\,\,\,\,\delta \ll
1,\,\,\Delta =0
\end{equation}
It is interesting to note that the same expressions (\ref{loc2}) and (\ref{loc1}) are 
obtained in Ref. \cite{EW-93} by assuming that localization length $l_\infty $ is 
determined by the reflection coefficient from a single
pair $\epsilon _n=\epsilon _{n+1}$ , embedded in a perfect lead. It is of
interest to check how accurate are estimates found in \cite{IKT-95} and \cite
{EW-93} if to apply them for any energy inside the band.

To find the localization lengths $l_N^{(1)}$ and $l_N^{(2)}$ for finite
samples of the size $N$, we have used the same approach described in the
previous section, by examining the eigenstates with energies in a small
energy window $\Delta E$ $\le 10^{-2}E_{cr}$ near the critical energy $%
E_{cr}=\epsilon _B$. Yet, since in the region of critical energies
$l_{\infty}(E)$ and thus the localization properties of eigenstates, depent 
from the energy in a singular way, (see Eq.~(\ref{loc2}) and Eq.~(\ref{loc1})) 
we took from the energy window only the eigenvector with
the corresponding eigenstate which is closer to $E_{cr}$ (but always
differant from it, $E \neq E_{cr}$). This was a natural choice in order to
study statistical properties of eigenstates with the similar
localization properties (i.e. the eigenstates just near the totally
extended one). The average values of $l_N^{(q)}$ have been obtained by
statistical averaging over an ensemble of more than 3000 samples of the size 
$N=100\div 800$ with different pair-correlated disorder. The results are
reported in Fig.~4 together with the fit to the expression (\ref{scalfq}).
One can see a quite good scaling of the form (\ref{scalfq}), in spite of
fluctuations which are much larger in this energy region compared to that in
the center of bands. The fitting coefficients $c_1=2.20$ and $c_2=1.06$ are
slightly less than those in the band center. This fact may be explained by
an approximate character of the analytical expression (\ref{our}) (one
should also note that for the values of $\beta _q$ very close to the limit $%
\beta _q=1$ the computational errors are very large).

In Fig.~5 the same data are represented in the variables (\ref{fit}), with
the fit correspondent to the dependence (\ref{YX}). It is interesting that
in spite of a slight difference for the coefficients $a_1=0.75$ and $%
a_2=0.08\,$ in comparison for those found in the center of bands, the shift $%
\Delta a_{1,2}=a_1-a_2=0.67$ remains almost the same (compared to 0.6).


\section{\bf Summary}


We have studied a 1D tight-binding model with binary on-site
disorder that is randomly assigned in every two sites. For such a model we
know that there exist two special energies $E_{cr}$ at which transparent
states appear \cite{DWP-90,F-89,B-92,IKT-95}. For other energies, but close
to critical ones, the localization length is very large, leading to
nearly-transparent states that are of great importance in the conducting
properties of finite samples. This property is quite different from genuine
disordered models of Anderson type.

Our numerical study of random dimer models of finite size deals with the
scaling properties of the eigenstates. This study was motivated by the
remarkable scaling law that has been found for different 1D and quasi-1D
models, both dynamical (Kicked Rotator on a torus \cite
{CGIS-90,I-90}) and disordered (1D Anderson and Lloyd models \cite
{CGIFM-92,M-93} and quasi-1D models \cite{FM-94,CMI-90}). These latter
results indicate that eigenstates in finite samples with disorder have
generic properties, regardless of the details of the disorder.

The main result of our computations is that scaling properties of
eigenstates of finite dimers are of the same type as for the disordered
models mentioned above in spite of the existence of nearly-transparent
states. In particular, both entropy localization length and localization
length from inverse participation ratio normalized in the proper way, follow
the universal scaling law of Eq. (\ref{scalfq}).

The scaling relation of Eq. (\ref{scalfq}) can be also represented in a very
intriguing form \cite{CGIFM-92,M-93,FM-92}: 
\begin{equation}
\label{fun}\frac 1{l_N^{(q)}(\epsilon ,E)}=\frac 1{l_\infty ^{(q)}(\epsilon
,E)}+\frac 1{l_N^{(q)}(0,E)} 
\end{equation}
which still has no physical ground. In Eq. (\ref{fun}), $l_N^{(q)}(\epsilon
,E)$ represents the localization length for a finite sample and finite
disorder, $l_\infty ^{(q)}(\epsilon ,E)$ is the localization length for an
infinite sample with the same disorder and $l_N^{(q)}(0,E)$ is the
localization length for a finite sample with zero disorder. One should
stress that all three localization lengths are defined in the same way given
through expressions of Eq. (\ref{lq}) and (\ref{Hq}). One can see that the
form Eq. (\ref{fun}) is parameter independent; the same form holds also for
the 1D Anderson and Lloyd models (see \cite{CGIFM-92,M-93}).

In our numerical study the energy window has been chosen in the middle of
the spectrum as well as close to the critical energy, giving the same
scaling form (\ref{scalfq}). The slight difference in the coefficients $%
c_q\, $ for these two energy regions seems to indicate that the analytical
expression (\ref{our}) needs some correction related to an additional
dependence on the energy when the latter is not close enough to the critical
one. Our results indicate that the same scaling is expected to hold for
other values of the energy inside the band. One should note that the scaling
(\ref{scalfq}) (or, the same, (\ref{fun})) can be used to check the accuracy
of expressions for the localization length $l_\infty $ in dependence of the
parameters $E$ and $\epsilon _b$ , if for some values of these parameters
the scaling function $\beta _q$ is found with a high accuracy.

It is of quite interest to check the scaling behaviour of localization 
lengths corresponding to the higher moments $q\ge 3$ in (\ref{lq}).
Analytical treatment \cite{FM-94} for disordered models have shown that
the scaling law (\ref{scalfq}) approximately holds also for higher moments.
The correct expression for $ \beta_q (\lambda) $ is known only in the limit
case of very localized $(\lambda \ll 1)$ and extended $(\lambda \gg
1)$ states. It has the same form (\ref{YX}) with $b_q=1$
but with different values of $a_q$ in these limits (see details
in \cite{FM-94}). On the base of our results for $q=1,2$, 
it is quite natural to expect that for the dimer model 
the correspondence to the
analytical predictions \cite{FM-94} also hold for higher moments, however, 
this question remains open.

Finally, we should comment that the results obtained in the present work can
generalized to cases with correlation blocks  larger than dimers, viz.
$m$-blocks with $m=3, 4, 5, ..$.  In these more general cases the following
surprizing result holds:  Given an arbitrary distribution of correlated
blocks with {\it even} length, i.e. an arbitrary distribution of dimers, 
quatromers, sextomers, octamers, etc, with the same energy $\epsilon$, that 
populate a lattice with sites that have some other energy value, there
is always a resonant energy $E_{cr} = \epsilon$ that corresponds to a delocalized
state.  This result can be easily deducted from the general expressions
of ref. \cite{IKT-95}.  On physical grounds, we expect the 
localization properties of the eigenstates of this system to
follow similar scaling laws to the ones derived in the present work.


\section{\bf Acknowledgements}

One of the authors (F.\thinspace M. I.) wishes to acknowledge the support of
Grant ERBCHRXCT\thinspace 930331 Human Capital and Mobility Network of the
European Community and also the support of Grant No RB7000 from the
International Science Foundation. We thank Paolo Grigolini Bruce J. West
and C. M. Soukoulis for discussions

\newpage

\centerline{\bf Figure Captions} \vspace{1.0cm}

\noindent {\bf Fig.1} Scaling of $\beta _q$ as a function of the
localization ratio $\lambda =l_\infty /N$ for the RDM with $Q=0.5$ and $%
\epsilon _A=0$. The energies $E$ are taken in an energy window centered at $%
E=\epsilon _B/2$ for values of $\epsilon
_B=2.0;\,1.8;\,1.6;\,1.2;\,1.0;\,0.8;\,0.6;\,0.4;\,0.35.$ Smooth curves
correspond to the dependence (\ref{scalfq}) with $c_q$ as a fitting
parameter. \thinspace

(a) Scaling for $\beta _1$ with $c_1=2.80$ ;

(b) Scaling for $\beta _2$ with $c_2=1.55.$\\

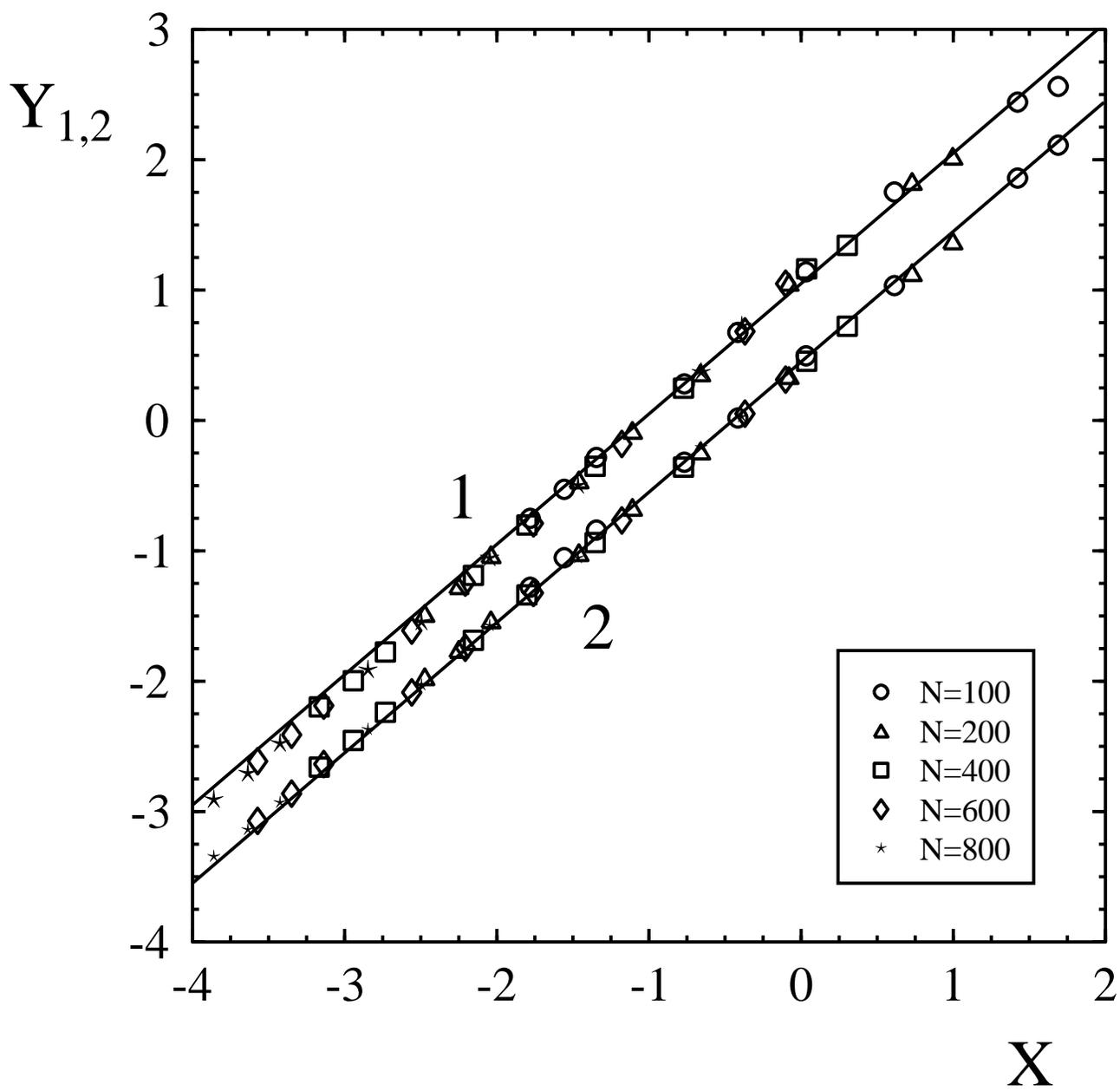
\noindent {\bf Fig.2} The scaling of $\beta _1$, $\beta _2$ as a function of 
$\lambda =l_\infty /N$ in the variables $Y_{1,2}$ and $X$ (see (\ref{fit}))
for the same values of the parameters as in Fig. 1. Straight lines (1) and
(2) correspond to the expression (\ref{YX}) with $a_1=1.05;\,\,b_1=1$ and $%
a_2=0.45;\,\,b_2=1$ respectively. \\

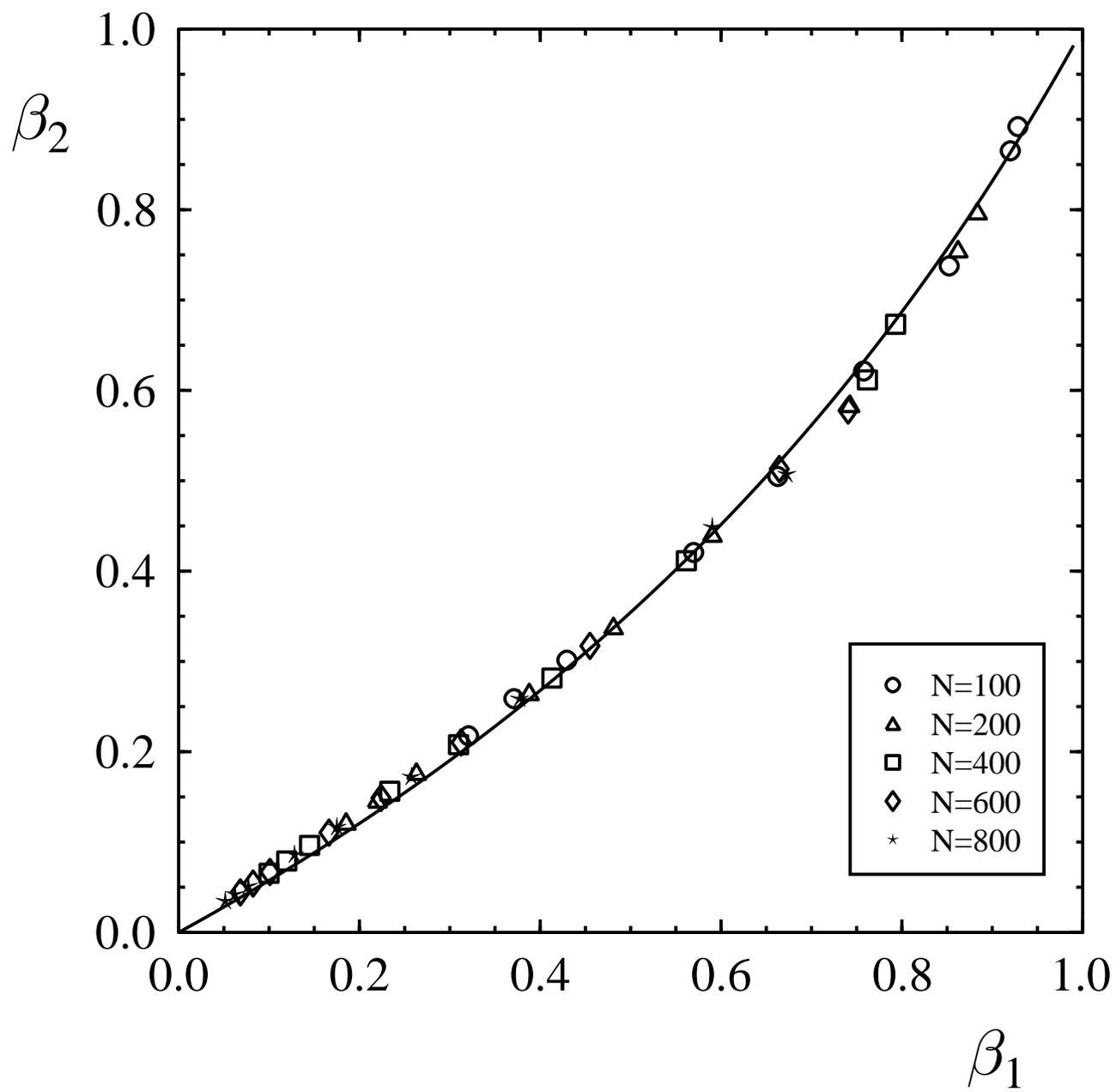
\noindent {\bf Fig.3} A plot of $\beta _2$ as a function of $\beta _1$. It
is interesting to note that the fitting curve has the same form (\ref{beta12}%
) as the ones for the case of $\beta _{1,2}$ plotted in Fig.1 as a function
of $\lambda =l_\infty /N$.\\

\noindent {\bf Fig.4} The same as in Fig. 1 for the energies close to the
critical one $(E_{cr}=\epsilon _B)$ , for $q=1$ (Fig.4a) and $q=2$ (Fig.4b).
The values of $\epsilon _B$ are taken as $\epsilon _B=1.8;\,1.6;\,1$%
.4;\thinspace 1.2; 1.0.\\

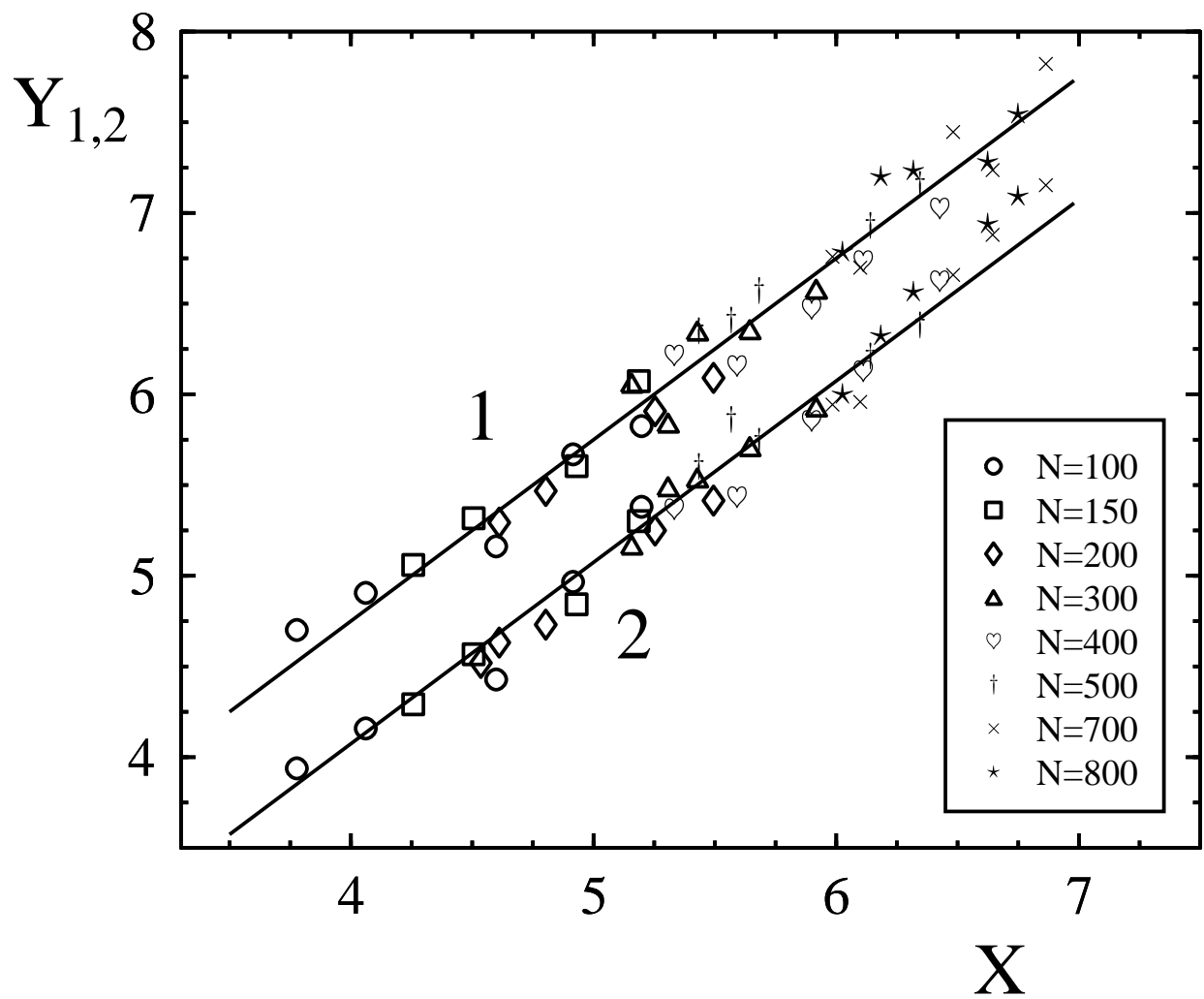
\noindent {\bf Fig.5} The same as in Fig.2 for the parameters of Fig.4.
Straight lines (1) and (2) correspond to the expression (\ref{YX}) with $%
a_1=0.75;\,\,b_1=1$ and $a_2=0.08;\,\,b_2=1$ respectively. \\


\begin{thebibliography}{99}
\bibitem{note} We have corrected the coefficient in Eq. (\ref{our}) by the factor 
$2$ comparing to that in Ref. \cite{IKT-95}.
\end{thebibliography}
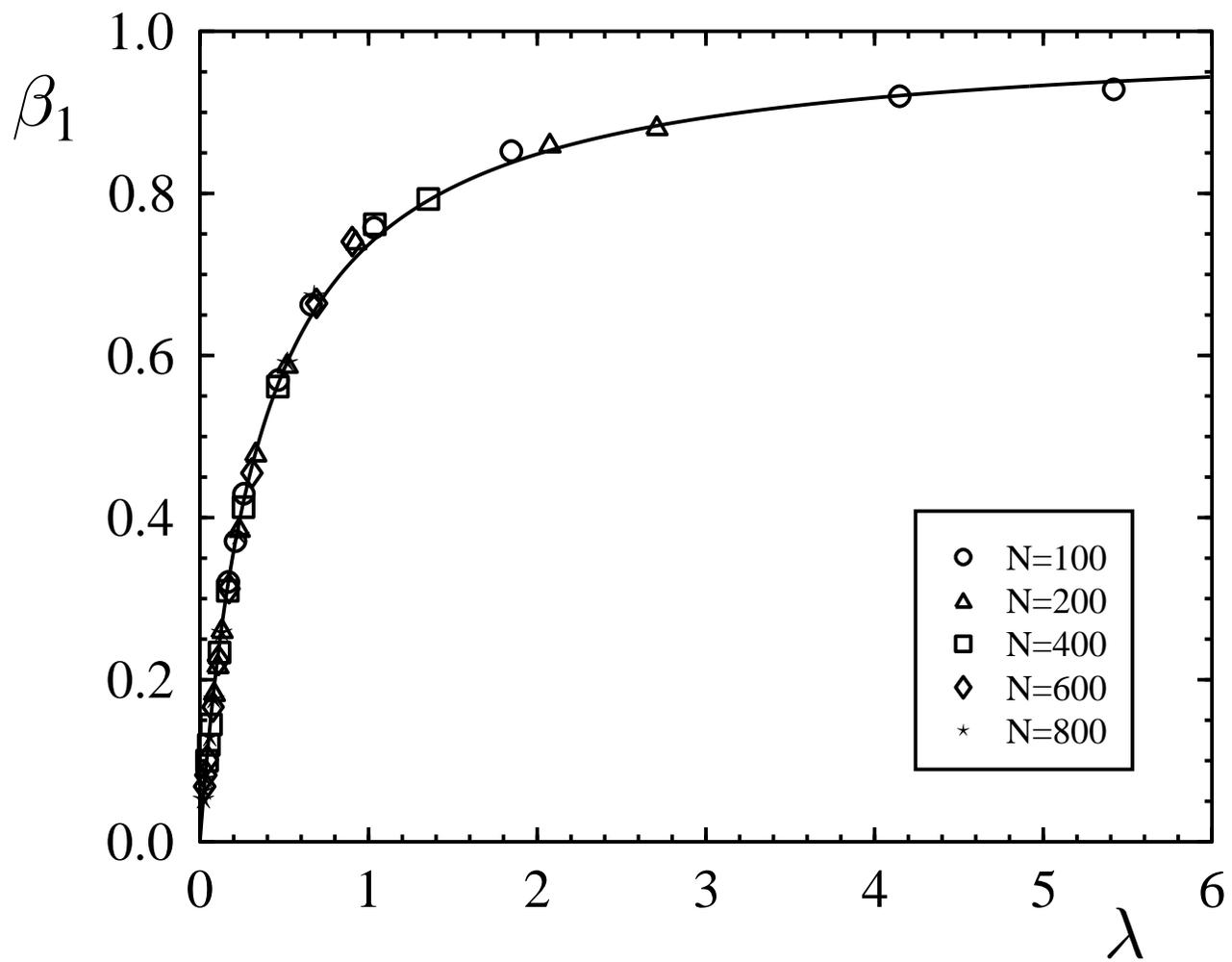
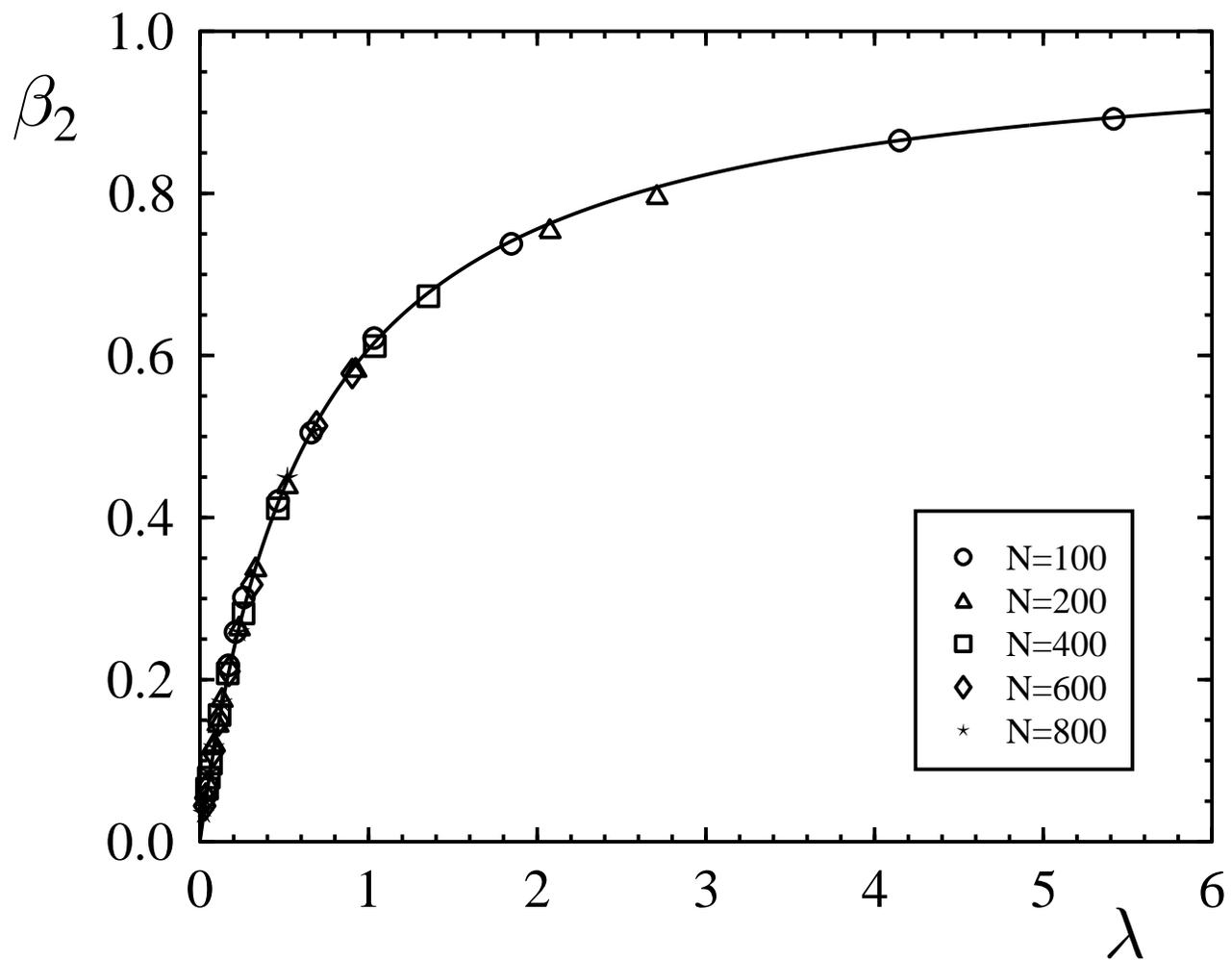
\end{document}